\documentclass{article}

\usepackage{arxiv}
\usepackage{graphicx}

\graphicspath{ {./} }
\usepackage[center]{caption}
\usepackage{csvsimple}

\usepackage[utf8]{inputenc} 
\usepackage[T1]{fontenc}    
\usepackage{hyperref}       
\usepackage{url}            
\usepackage{booktabs}       
\usepackage{amsfonts}       
\usepackage{nicefrac}       
\usepackage{microtype}      
\usepackage{lipsum}

\usepackage{authblk}

\author[1,*]{Justin D Krogue MD}
\author[4]{Kaiyang Cheng (Victor)}
\author[1]{Kevin M Hwang MD}
\author[1]{Paul Toogood MD}
\author[1]{Eric G Meinberg MD1}
\author[1]{Erik J Geiger MD}
\author[1]{Musa Zaid MD}
\author[3]{Kevin C McGill MD MPH}
\author[3]{Rina Patel MD}
\author[3]{Jae Ho Sohn MD MS}
\author[3]{Alexandra Wright MD}
\author[2]{Bryan F Darger MD}
\author[2]{Kevin A Padrez MD}
\author[3]{Eugene Ozhinsky PhD}
\author[3]{Sharmila Majumdar PhD}
\author[3]{Valentina Pedoia PhD}

\affil[1]{Department of Orthopaedic Surgery, University of California San Francisco}
\affil[2]{Department of Emergency Medicine, University of California San Francisco}
\affil[3]{Department of Radiology and Biomedical Imaging, University of California San Francisco}
\affil[4]{University of California, Berkeley}

\affil[*]{Corresponding Author: Justin D Krogue, justin.d.krogue@gmail.com}

\title{Automatic Hip Fracture Identification and Functional Subclassification with Deep Learning}

\begin{document}
\maketitle

\begin{abstract}
\textbf{Purpose}: Hip fractures are a common cause of morbidity and mortality. Automatic identification and classification of hip fractures using deep learning may improve outcomes by reducing diagnostic errors and decreasing time to operation.
\textbf{Methods}: Hip and pelvic radiographs from 1118 studies were reviewed and 3034 hips were labeled via bounding boxes and classified as normal, displaced femoral neck fracture, nondisplaced femoral neck fracture, intertrochanteric fracture, previous ORIF, or previous arthroplasty. A deep learning-based object detection model was trained to automate the placement of the bounding boxes. A Densely Connected Convolutional Neural Network (DenseNet) was trained on a subset of the bounding box images, and its performance evaluated on a held out test set and by comparison on a 100-image subset to two groups of human observers: fellowship-trained radiologists and orthopaedists, and senior residents in emergency medicine, radiology, and orthopaedics.
\textbf{Results}: The binary accuracy for fracture of our model was 93.8\% (95\% CI, 91.3-95.8\%), with sensitivity of 92.7\% (95\% CI, 88.7-95.6\%), and specificity 95.0\% (95\% CI, 91.5-97.3\%). Multiclass classification accuracy was 90.4\% (95\% CI, 87.4-92.9\%). When compared to human observers, our model achieved at least expert-level classification under all conditions. Additionally, when the model was used as an aid, human performance improved, with aided resident performance approximating unaided fellowship-trained expert performance.
\textbf{Conclusions}: Our deep learning model identified and classified hip fractures with at least expert-level accuracy, and when used as an aid improved human performance, with aided resident performance approximating that of unaided fellowship-trained attendings.

\end{abstract}

\section{Introduction}
Hip fractures are a significant cause of morbidity and mortality in the United States and throughout the world, with more than 300,000 occurring in 2014 in the United States alone (1). Although age-adjusted hip fracture incidence has decreased in recent years, absolute numbers of hip fracture are expected to increase by 12\% by 2030 owing to an aging population (2). Hip fractures, especially in elderly patients, represent a life-changing event, and carry a significant risk of decreased functional status and death, with one-year mortality rates reported to be as high as 30\% (3,4). 

Accurate and timely diagnosis of hip fractures is critical, as outcomes are well-known to depend on time to operative intervention (5,6,7). Specifically, Maheshwari et al recently showed that each 10-hour delay from admission to surgery is linearly associated with a 5\% higher odds of 1-year mortality  (6). Efficient radiographic identification and classification of hip fracture represents a key component to optimizing outcomes by avoiding unnecessary delays, especially as implant choice for a hip fracture depends almost entirely on its radiographic classification, and the initial image often contains enough information to begin planning the definitive surgery (Figure 1) (8,9).

\begin{figure}
  \centering
  \includegraphics{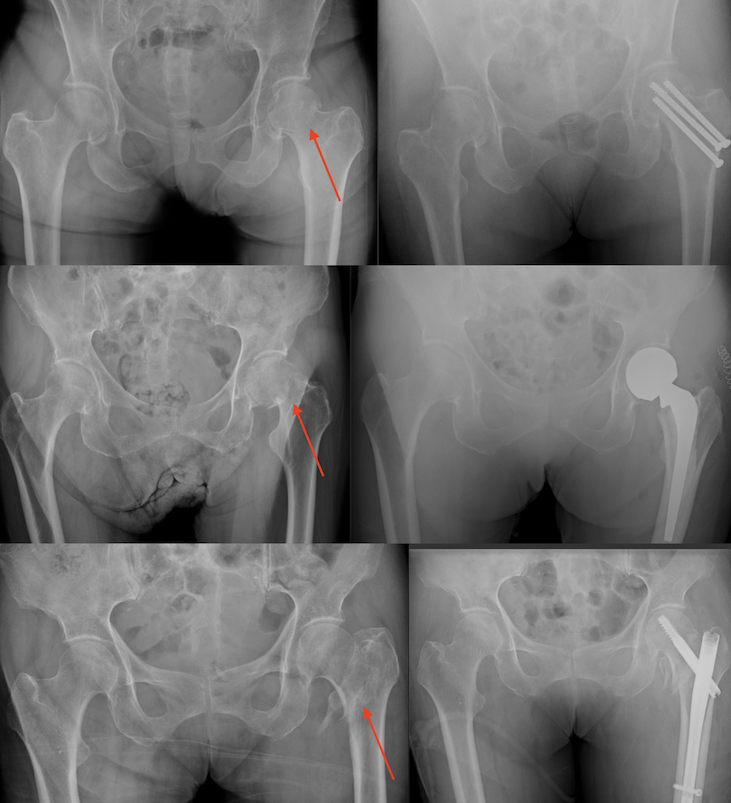}
  \caption{Implant choice by fracture type. Top row: non-displaced femoral neck fracture, which is treated with cannulated screw fixation. Second row: displaced femoral neck fracture, treated with arthroplasty. Third row: intertrochanteric fracture, which is treated with internal fixation with cephalomedullary nail. Red arrows point to fractures.}
  \label{fig:fig1}
\end{figure}

Machine learning, and deep learning with artificial neural networks in particular, have recently shown great promise in achieving human or near-human level performance in a variety of highly complex perceptual tasks traditionally challenging for machines to perform, including image classification and natural language processing. Artificial neural networks exploit a stacked architecture of layers of “neurons” to learn hierarchical representations of data across multiple levels of abstraction, calculating more and more complex features in each layer. Convolutional neural networks, the standard in computer vision, use sets of filters in each layer to generate many complex features from an input image and have shown great promise in many areas of radiography, including in many musculoskeletal applications (10,11,12,13,14,15,16).

In this study, we propose an automated system of hip fracture diagnosis and classification using deep learning with a convolutional neural network. Such a system has enormous clinical importance as it may decrease the rate of missed fractures and the time to operative intervention, thus potentially improving patient outcomes. We hypothesize that this system will be at least equivalent to expert performance in hip fracture identification and classification and will improve physician performance when its predictions are used as an aid.

\section{Materials and Methods}
\subsection{Dataset Acquisition}
After obtaining IRB approval, our radiology report database was queried retrospectively for hip/pelvic radiographs obtained in the emergency room with the words “intertrochanteric” or “femoral neck” occurring near “fracture” from 1998-2017 in patients >= 18 years old. 919 of these studies were identified as likely containing a hip fracture based on manual review of the reports and included in the study. An additional 199 studies were chosen at random from the database of hip/pelvic radiographs using the same year and age cutoffs. Each image from these 1118 studies was then extracted and processed using the Python Pydicom package (version 1.1.0). Subject flowchart is shown in Figure 2.
 
\begin{figure}
  \centering
  \includegraphics{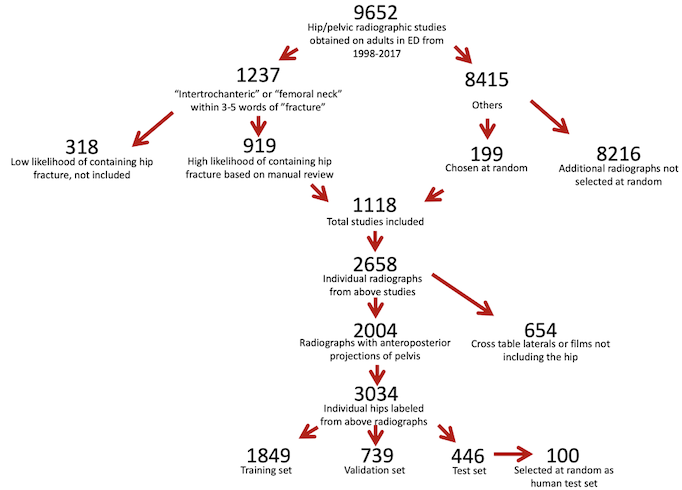}
  \caption{subject flowchart. From 9652 eligible radiographs, 919 were identified as having high probability of hip fracture from the use of regular expressions and manual review of reports. An additional 199 were included at random. The 2658 radiographs from these 1118 studies were then extracted and manually reviewed, and all 2004 radiographs containing a hip and an anteroposterior projection of the pelvis were included. From these radiographs, 3034 individual hips were identified and labeled, and these images were divided into training, validation and test sets using a 60:25:15 split. 100 images were chosen at random from the test set to be used for comparison to human performance.}
  \label{fig:fig2}
\end{figure}

All images were reviewed by two postgraduate-level 4 (PGY-4) orthopaedic residents (J.K., K.H.), using the Visual Geometry Group Image Annotator (University of Oxford, Oxford, United Kingdom) (17). All anteroposterior (AP) projections of the pelvis were included; cross-table lateral views and images not including the hip were excluded. Bounding boxes were drawn around each hip, and each was classified as unfractured, fractured, or containing hardware. Fractures were further subclassified as either nondisplaced femoral neck (FN) fractures, displaced FN fractures, or intertrochanteric (IT) fractures. Hardware was subclassified as previous internal fixation (ORIF) or arthroplasty, and is counted as “no fracture” in binary fracture prediction. In cases of uncertainty, the patient’s subsequent imaging was reviewed and further CT, MRI, and post-operative imaging were used as ground truth. If an operation eventually occurred, the label was inferred from the operative fixation chosen (Figure 1). 3034 hips were labeled in this fashion, and were split by accession number into training, validation, and test sets using a 60:25:15 split, with randomization by class distribution to ensure an equal distribution of classes between datasets. In this way all images from a study appeared in only one dataset.

\subsection{Data Processing and Augmentation}
Prior to insertion in the model, the hip images were resized to 224x224 pixels and replicated into three channels in order to be compatible with the ImageNet-pretrained model, and left hips were flipped to appear as right hips. To make our model invariant to differences in zoom of the bounding box, each hip in the training set appeared twice, with differing sizes of bounding boxes. To each of these images, we applied data augmentation with 3 types of contrast changing, cutout (18), gaussian-mixture masking, and bounding box wiggling to generate 6 additional images (Figure 3).

\begin{figure}
  \centering
  \includegraphics{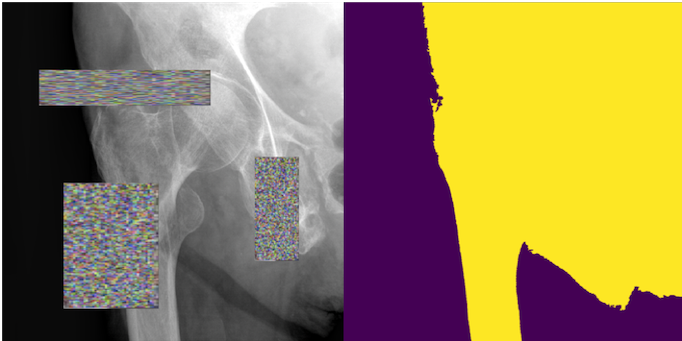}
  \caption{cutout and Gaussian mixture masking applied to single image. In cutout (left), random areas of the image are cut out to provide regularization during training, which can be described as image-level dropout. Gaussian mixture masking (right) takes the histogram of the image and segments out the most active area, which in most cases includes the bony structures. In the small number of cases in which some of the hip itself is masked out, the image serves as a useful regularization sample during training similar to cutout.}
  \label{fig:fig3}
\end{figure}

\subsection{Model Architecture}
A Densely Connected Convolutional Neural Network (DenseNet) architecture consisting of 169 layers was chosen for fracture classification. In a DenseNet convolutional layers are placed in discrete “dense blocks,” and within those blocks a layer receives as input all activations from the previous layers within the block (19). We added an attention pooling mechanism after the final dense block, which acts as a learnable weight mask over the image, allowing prioritization of the most important features. The final layer is a softmax layer with one output for every hip class (Figure 4).

\begin{figure}
  \centering
  \includegraphics{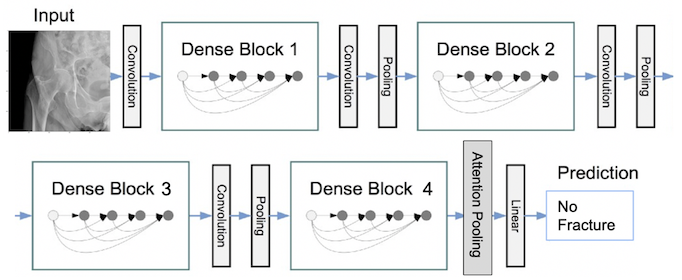}
  \caption{model architecture (figure modified from Huang and Liu et al (19)). Four dense blocks are stacked with intervening “transition layers” which use convolutions and pooling to shrink feature-map sizes between dense blocks. In a dense block each individual layer receives as an input all activations from previous layers in that block. In this way feature reuse is encouraged, resulting in a more compact model with less overfitting. In a modification to the original Densenet architecture, an attention pooling layer is added after the final dense block, which uses learned weights to pool the feature map with minimal loss of important features. This is fed into a densely-connected layer and finally to a softmax layer which outputs a prediction.}
  \label{fig:fig4}
\end{figure}

\subsection{Model Training}
The DenseNet was initialized with ImageNet pretrained weights (20) and trained using Adam (21) with learning rate of 0.00001, batch size of 25, and learning rate decay of 0.9. Training was stopped after 10 epochs passed without improvement in validation accuracy, and the model with the highest validation set accuracy was then chosen. At inference time the binary prediction is computed by summing the probabilities of the fractured and the unfractured classes.

\subsection{Bounding Box Detection}
In order to automate the process of hip fracture detection end-to-end, it is necessary to train an object detection algorithm to place the bounding boxes automatically. The object detection network was implemented in Python with TensorFlow Object Detection API (Google, Mountain View, CA) on a single-shot detector with Resnet-50 feature pyramid network architecture (22, 23). The model’s output consisted of bounding boxes around the upper extremity of the femur and labels of left vs right hip. Non-max suppression was performed to eliminate redundant boxes with constraints of no more than one box per class in a given image and intersection over union threshold of 0.3. The input data was augmented by randomly cropping the images. The model was pre-trained on ImageNet classification and COCO object detection datasets and trained with Nvidia Titan X GPU for 25000 iterations (347 epochs, batch size: 16 images) on the training dataset of radiographs with bounding boxes defined by a PGY-4 orthopaedic resident. To evaluate the performance of the network, inference has been performed on the radiographs from the validation set, and finally, on the test set. Detection accuracy was measured with the intersection-over-union metric, and the performance of the DenseNet classification algorithm was compared using manually-detected vs automatically-detected bounding boxes.

\subsection{Model Evaluation and Statistical Analysis}
The trained model’s performance was evaluated using the receiver-operator curve (ROC) and its area under the curve (AUC), and via calculation of key performance metrics including accuracy, sensitivity, and specificity, with 95\% confidence intervals calculated via Jeffrey’s prior interval for a binomial distribution.

100 images were chosen at random from the test set for comparison to human evaluators. As our human experts, we selected two trauma-fellowship trained orthopaedic surgeons (P.T. and E.G.M., average 10 years post-fellowship) and two musculoskeletal (MSK) fellowship-trained radiologists (K.C.M. and R.P., average 2 years post-fellowship). As residents often perform the initial film interpretation in an academic setting, two PGY-4 residents in each of the fields of emergency medicine, orthopaedics, and radiology were also selected (A.W., J.H.S., K.A.P., B.F.D., E.J.G., M.Z). Each physician was shown the 100 images exactly as input into the model (“model-quality” images), and after one week, they evaluated the same hips in shuffled order at the full resolution and size (“full-quality” images). In order to assess the effect of model-aided image reading, each physician was finally presented with the model’s heatmap and top two suggestions when their answer differed from the model’s in the full-quality images, and they were asked to provide a final prediction (Figure 5).

\begin{figure}
  \centering
  \includegraphics{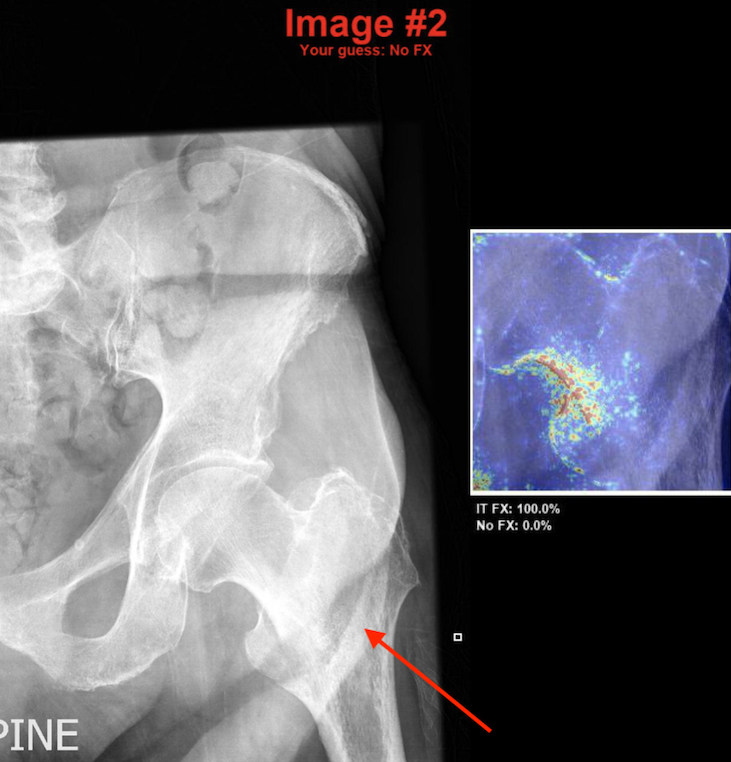}
  \caption{model-aided conditions. In cases where the human observer’s answer differed from the model, they were shown the original image with their prediction along with the model heatmap and top two model predictions with probabilities. In this case the human observer is presented with the model’s prediction of IT fracture (denoted by the red arrow added manually for the purpose of this figure), which is correct, after stating that there was no fracture.}
  \label{fig:fig5}
\end{figure}

Key performance metrics were calculated for each group of observers, and Cohen’s kappa coefficients were then calculated to measure each observer’s agreement with the ground truth. 95\% confidence intervals for kappa coefficients were calculated by sampling with replacement for 10000 iterations, and the differences in Cohen’s Kappa coefficients were compared via a randomization test with 10000 permutations and a significance value of p < 0.05.

\section{Results}
\subsection{Model Performance}
The average patient age included in the study was 74.6 years old (standard deviation 17.3), with 62\% females. The age, sex, multiclass and binary class distributions with ages of our data set are shown in Table 1. Using Pearson’s chi-squared test, there was no statistically significant difference in distribution between the different datasets (p-value 0.886 for multiclass distributions, p-value 0.897 for binary distributions).

\begin{table}[]
\centering
\caption{age, sex, multiclass and binary distribution amongst the radiographs examined}
\label{tab:table1}
\resizebox{\textwidth}{!}{%
\begin{tabular}{l|lllll}
\hline
-                         & Overall, n=3034    & Training, n=1849   & Validation, n=739  & Test, n=446        & Human test, n=100  \\
\hline
Age (SD)                  & 74.3 (+/- 17.6)    & 74.5 (+/- 17.3)    & 75.2 (+/- 16.5)    & 76.6 (+/- 17.3)    & 78.3 (+/- 12.8)    \\
Sex                       & 62.9\% F, 37.1\% M & 60.8\% F, 39.1\% M & 64.9\% F, 35.1\% M & 58.1\% F, 41.9\% M & 61.5\% F, 38.5\% M \\
No fracture               & 1327 (43.7\%)      & 815 (44.1\%)       & 326 (44.1\%)       & 186 (41.7\%)       & 42 (42\%)          \\
IT fracture               & 766 (25.2\%)       & 458 (24.7\%)       & 187 (25.3\%)       & 121 (27.1\%)       & 27 (27\%)          \\
FN fracture, displaced    & 527 (17.4\%)       & 315 (17.0\%)       & 138 (18.7\%)       & 74 (16.6\%)        & 17 (17\%)          \\
FN fracture, nondisplaced & 183 (6.0\%)        & 113 (6.1\%)        & 43 (5.8\%)         & 27 (6.1\%)         & 6 (6\%)            \\
Arthroplasty              & 172 (5.7\%)        & 113 (6.1\%)        & 27 (3.7\%)         & 32 (7.2\%)         & 7 (7\%)            \\
ORIF                      & 59 (1.9\%)         & 35 (1.9\%)         & 18 (2.4\%)         & 6 (1.3\%)          & 1 (1\%)            \\
Total: Unfractured        & 1558 (51.4\%)      & 963 (52.1\%)       & 371 (50.2\%)       & 224 (50.2\%)       & 50 (50\%)          \\
Total: Fractured          & 1476 (48.6\%)      & 886 (47.9\%)       & 368 (49.8\%)       & 222 (49.8\%)       & 50 (50\%)         
\end{tabular}}
\end{table}

When evaluated on the overall held out test set, the model’s binary accuracy for the presence of fracture is 93.8\% (95\% CI, 91.3-95.8\%), with sensitivity 92.7\% (95\% CI, 88.7-95.6\%), and specificity 95.0\% (95\% CI, 91.5-97.3\%). Multiclass accuracy is 90.4\% (95\% CI, 87.4-92.9\%), with sensitivities, and specificities for each class type shown in Table 2 and a confusion matrix shown in Table 3. Specificity was universally high for all fracture types (>= 96.9\%), indicating very few false positive diagnoses. While sensitivity for displaced FN fractures was 86\%, 100\% of these were classified as a fracture of some type, indicating 100\% binary sensitivity for these fracture types. Similarly, while approximately half of nondisplaced FN fractures were correctly identified as such, nearly 60\% were identified as FN fractures of some type. An ablation table showing the effect on multiclass accuracy over the validation and test sets of our image augmentation techniques and attention mechanism is shown in Table 4.

\begin{table}[]
\centering
\caption{multiclass performance metrics of the CNN regarding each classification subtype}
\label{tab:table2}
\begin{tabular}{l|lllll}
\hline
Category                  & Sensitivity, \% (95\% CI) & Specificity, \% (95\% CI) &  &  &  \\
\hline
No fracture               & 94.5 (90.5-97.1)          & 92.6 (88.9-95.3)          &  &  &  \\
IT fracture               & 93.3 (87.8-96.8)          & 96.9 (94.5-98.4)          &  &  &  \\
FN fracture, displaced    & 87.5 (78.4-93.6)          & 98.9 (97.4-99.6)          &  &  &  \\
FN fracture, nondisplaced & 46.2 (28.2-64.9)          & 97.8 (96.1-98.9)          &  &  &  \\
Arthroplasty              & 96.9 (86.3-99.7)          & 100 (99.4-100.0)          &  &  &  \\
ORIF                      & 100 (67.0-100.0)          & 100 (99.4-100.0)          &  &  &  \\
\end{tabular}
\end{table}

\begin{table}[]
\centering
\caption{normalized confusion matrix of multiclass classification. Row headings represent the true label, and column headings represent the model’s prediction.}
\label{tab:table3}
\resizebox{\textwidth}{!}{%
\begin{tabular}{l|llllll}
\hline
                          & No fracture & IT fracture & \begin{tabular}[c]{@{}l@{}}FN fracture,\\ displaced\end{tabular} & \begin{tabular}[c]{@{}l@{}}FN fracture,\\ nondisplaced\end{tabular} & Arthroplasty & ORIF  \\
\hline
No fracture               & 95\%        & 3\%         & 1\%                                                              & 2\%                                                                 & 0\%          & 0\%   \\
IT fracture               & 6\%         & 93\%        & 0\%                                                              & 1\%                                                                 & 0\%          & 0\%   \\
FN fracture, displaced    & 0\%         & 7\%         & 86\%                                                             & 7\%                                                                 & 0\%          & 0\%   \\
FN fracture, nondisplaced & 42\%        & 0\%         & 12\%                                                             & 46\%                                                                & 0\%          & 0\%   \\
Arthroplasty              & 3\%         & 0\%         & 0\%                                                              & 0\%                                                                 & 97\%         & 0\%   \\
ORIF                      & 0\%         & 0\%         & 0\%                                                              & 0\%                                                                 & 0\%          & 100\%
\end{tabular}}
\end{table}

\begin{table}[]
\centering
\caption{ablation table showing effect of image augmentation and attention mechanism on validation and test set multiclass accuracies with 95\% confidence intervals. Final model is bolded.}
\label{tab:table4}
\resizebox{\textwidth}{!}{%
\begin{tabular}{l|llllll}
\hline
                                                                                                                                                     & \begin{tabular}[c]{@{}l@{}}Validation set accuracy\\ \% (95\% CI)\end{tabular} & \begin{tabular}[c]{@{}l@{}}Test set accuracy\\ \% (95\% CI)\end{tabular} &  &  &  &  \\
\hline
Without metadata preprocessing                                                                                                                       & 83.4 (80.6-85.9)                                                               & 81.4 (77.6-84.8)                                                         &  &  &  &  \\
\hline
With preprocessing                                                                                                                                   & 81.2 (78.3-83.9)                                                               & 81.5 (77.7-84.9)                                                         &  &  &  &  \\
\hline
With attention mechanism                                                                                                                             & 84.1 (81.3-86.6)                                                               & 81.3 (77.5-84.7)                                                         &  &  &  &  \\
\hline
With Gaussian mixture                                                                                                                                & 84.7 (81.9-87.1)                                                               & 82.7 (78.9-85.9)                                                         &  &  &  &  \\
\hline
With two bounding box crops and wiggling                                                                                                             & 91.5 (89.3-93.3)                                                               & 86.8 (83.4-89.7)                                                         &  &  &  &  \\
\hline
With all augmentations                                                                                                                               & 93.2 (91.2-94.9)                                                               & 89.5 (86.4-92.1)                                                         &  &  &  &  \\
\hline
\textbf{With attention mechanism and all augmentations}                                                                                              & \textbf{93.4 (91.4-95.0)}                                                      & \textbf{90.4 (87.4-92.9)}                                                &  &  &  &  \\
\hline
\begin{tabular}[c]{@{}l@{}}With attention mechanism and all augmentations,\\ hips flipped as part of augmentation and not preprocessing\end{tabular} & 93.1 (91.1-94.8)                                                               & 89.0 (86.6-91.1)                                                         &  &  &  &  \\
\hline
\begin{tabular}[c]{@{}l@{}}With attention mechanism and all augmentations\\ except Gaussian mixture\end{tabular}                                     & 92.6 (90.5-94.3)                                                               & 89.0 (85.8-91.7)                                                         &  &  &  &  \\
\hline
\begin{tabular}[c]{@{}l@{}}With attention mechanism and all augmentations\\ without metadata preprocessing\end{tabular}                              & 93.0 (90.9-94.6)                                                               & 89.5 (86.4-92.1)                                                         &  &  &  & 
\end{tabular}}
\end{table}

Cohen’s kappa coefficient for binary classification is .877 (95\% CI .830-.918) and for multiclass classification is .862 (95\% CI .822-.901). Binary classification ROC has AUC of 0.973, indicating excellent agreement with the ground truth, and is shown with multiclass ROCs and respective AUCs for each class type in Figure 6. AUCs generally were near one, indicating excellent agreement with the ground truth, with somewhat lower performance for nondisplaced FN fractures with AUC of 0.868.

\begin{figure}
  \centering
  \includegraphics{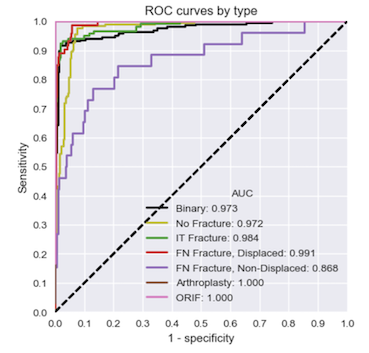}
  \caption{model’s receiver-operator curves (ROC) for each classification subtype. Binary represents the model’s ROC for detecting hip fracture overall.}
  \label{fig:fig6}
\end{figure}

Heatmaps for correctly-predicted images in each of the six categories are shown in Figure 7. Qualitative assessment of these images indicates high importance to cortical outlines in fracture classification, while the lucency of the fracture line itself appears to receive comparatively little attention.

\begin{figure}
  \centering
  \includegraphics{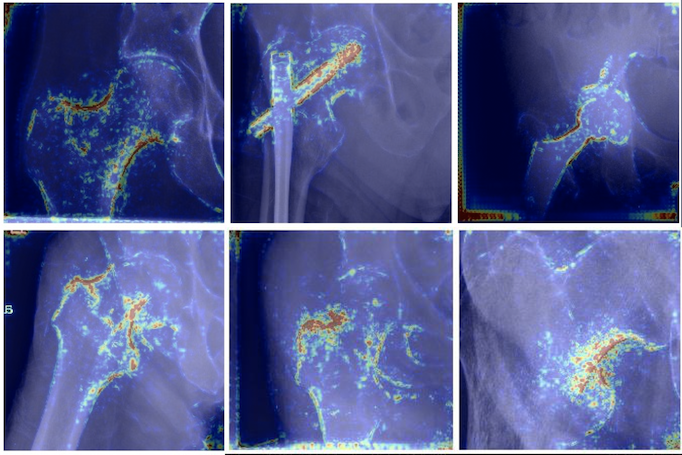}
  \caption{examples of heatmaps for the model’s correct predictions for each of the six classification types (from top left clockwise: no fracture, ORIF, arthroplasty, IT fracture, nondisplaced FN fracture, and displaced FN fracture). Of note, the model appears to pay attention to cortical outlines to make its classification, while the lucent fracture line appears to receive very little attention.}
  \label{fig:fig7}
\end{figure}

\subsection{Bounding Box Detection}
The trained RetinaNet object detection algorithm correctly identified every labeled hip in the test dataset with average intersection-over-union value of 0.92 (standard deviation 0.04, minimum value 0.64). In six radiographs, the detection algorithm labeled a hip that had not been labeled by the evaluator as it was only partially contained in the image. An example radiograph with manual and automatically-labeled boxes is shown in Figure 8. The DenseNet achieved 93.4\% binary accuracy (95\% CI, 90.8-95.4\%) and multiclass accuracy of 90.4\% (95\% CI, 87.4-92.9\%) on the automatically-generated bounding boxes, which did not differ significantly from the performance on manually-labeled boxes as measured by the difference in Cohen’s kappa for binary classification (p-value 0.779) and multiclass classification (p-value 0.9786). These results are shown in more detail in Table 5. 
 
\begin{table}[]
\centering
\caption{cohen’s kappa values for binary and multiclass performance of the DenseNet on the test set when using manually-labeled bounding boxes, and when using automatically-generated bounding boxes}
\label{tab:table5}
\begin{tabular}{l|lllll}
\hline
                           & Manual bounding box   & Automatic bounding box & \begin{tabular}[c]{@{}l@{}}Difference in kappa values\\ (two-tailed p-value)\end{tabular} &  &  \\
\hline
Binary kappa (95\% CI)     & 0.877 (0.831 - 0.918) & 0.868 (0.817 - 0.913)  & 0.009 (0.779)                                                                             &  &  \\
Multiclass kappa (95\% CI) & 0.865 (0.824 - 0.902) & 0.864 (0.824 - 0.901)  & 0.001 (0.979)                                                                             &  &  \\
\end{tabular}
\end{table}

\begin{figure}
  \centering
  \includegraphics{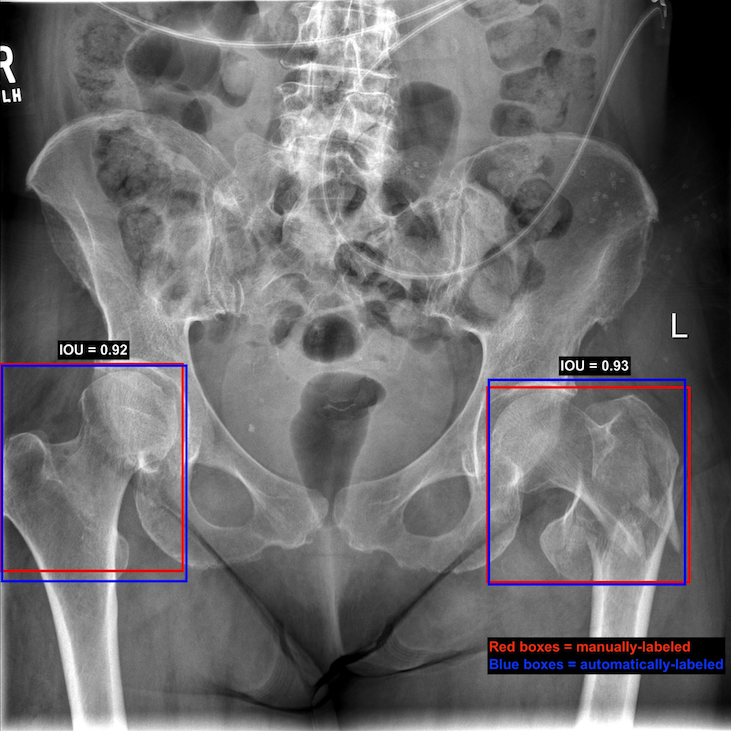}
  \caption{manual versus automated bounding box placement on an image from our test set.  In this image red boxes represent the manually-labeled boxes, while the blue boxes are the output of the box detection model. Intersection-over-union of the right hip is 0.92, and for the left hip is 0.93.  The right hip is not fractured here, while the left hip has an intertrochanteric fracture.}
  \label{fig:fig8}
\end{figure}

\subsection{Comparison to Human Performance}
Results of the human interpretation vs model performance of the 100-image subset are shown in Table 6, and the sensitivities/specificities of the pooled experts and residents with 95\% CIs are plotted on the model’s ROC in Figure 9. Performance of the human observers for each of the fracture subtypes is shown in Table 7. As validation of the ground truth, all labels were found to match the consensus expert predictions in the 78 cases in which all experts’ predictions agree.
 
\begin{table}[]
\centering
\caption{performance metrics of the CNN vs human observers in 100-image test subset}
\label{tab:table6}
\resizebox{\textwidth}{!}{%
\begin{tabular}{l|llllll}
\hline
                                   & \begin{tabular}[c]{@{}l@{}}Binary accuracy, \%\\ (95\% CI)\end{tabular} & \begin{tabular}[c]{@{}l@{}}Binary sensitivity, \%\\ (95\% CI)\end{tabular} & \begin{tabular}[c]{@{}l@{}}Binary specificity, \%\\ (95\% CI)\end{tabular} & \begin{tabular}[c]{@{}l@{}}Multiclass accuracy, \%\\ (95\% CI)\end{tabular} & \begin{tabular}[c]{@{}l@{}}Binary Cohen’s\\ kappa, (95\% CI)\end{tabular} & \begin{tabular}[c]{@{}l@{}}Multiclass Cohen’s\\ kappa, (95\% CI)\end{tabular} \\
\hline
Model                              & 96.0 (92.6-98.1)                                                        & 100.0 (97.5-100.0)                                                         & 92.0 (85.5-96.1)                                                           & 93.0 (88.8-95.9)                                                            & .920 (.838-.980)                                                          & .903 (.827-.960)                                                              \\
Experts, model-quality images      & 89.0 (85.7-91.8)                                                        & 95.5 (91.9-97.8)                                                           & 82.5 (76.8-87.3)                                                           & 83.5 (79.6-86.9)                                                            & .780 (.718-.840)                                                          & .775 (.725-.822)                                                              \\
Experts, full-quality images       & 93.8 (91.1-95.8)                                                        & 92.5 (88.2-95.6)                                                           & 95.0 (91.3-97.4)                                                           & 90.2 (87.1-92.9)                                                            & .875 (.825-.920)                                                          & .863 (.821-.901)                                                              \\
Experts, model-aided performance   & 95.8 (93.4-97.4)                                                        & 95.5 (91.9-97.8)                                                           & 96.0 (92.6-98.1)                                                           & 93.0 (90.2-95.2)                                                            & .915 (.874-.950)                                                          & .902 (.867-.934)                                                              \\
Residents, model-quality images    & 84.8 (81.8-87.5)                                                        & 91.0 (87.4-93.8)                                                           & 78.7 (73.8-83.0)                                                           & 76.7 (73.2-79.9)                                                            & .697 (.638-.753)                                                          & .685 (.639-.729)                                                              \\
Residents, full-quality images     & 86.5 (83.6-89.1)                                                        & 95.7 (92.9-97.5)                                                           & 77.3 (72.3-81.8)                                                           & 79.3 (76.0-82.4)                                                            & .730 (.675-.782)                                                          & .723 (.681-.764)                                                              \\
Residents, model-aided performance & 91.5 (89.1-93.5)                                                        & 98.0 (95.9-99.2)                                                           & 85.0 (80.6-88.7)                                                           & 88.7 (85.9-91.0)                                                            & .830 (.785-.873)                                                          & .846 (.812-.879)                                                              \\
\end{tabular}}
\end{table}

\begin{table}[]
\centering
\caption{sensitivity and specificity values for each fracture subtype for the model and human observers under all conditions on the 100-image human test set}
\label{tab:table7}
\resizebox{\textwidth}{!}{%
\begin{tabular}{l|ll|ll|ll}
\hline
                                   & IT fracture               &                           & FN fracture, displaced    &                           & FN fracture, nondisplaced &                           \\
\hline
                                   & Sensitivity, \% (95\% CI) & Specificity, \% (95\% CI) & Sensitivity, \% (95\% CI) & Specificity, \% (95\% CI) & Sensitivity, \% (95\% CI) & Specificity, \% (95\% CI) \\
\hline
Model                              & 100.0 (91.2-100.0)        & 100.0 (96.6-100.0)        & 88.2 (67.3-97.5)          & 98.8 (94.5-99.9)          & 66.7 (28.6-92.3)          & 94.7 (88.7-97.9)          \\
Experts, model-quality images      & 93.5 (87.7-97.1)          & 96.6 (94.0-98.2)          & 86.8 (77.2-93.2)          & 95.8 (93.2-97.6)          & 37.5 (20.4-57.4)          & 91.2 (88.0-93.8)          \\
Experts, full-quality images       & 91.7 (85.3-95.8)          & 98.3 (96.3-99.3)          & 88.2 (79.0-94.3)          & 98.2 (96.3-99.2)          & 50.0 (31.0-69.0)          & 96.5 (94.3-98.0)          \\
Experts, model-aided performance   & 96.3 (91.4-98.7)          & 99.7 (98.4-100.0)         & 89.7 (80.9-95.3)          & 98.5 (96.7-99.4)          & 62.5 (42.6-79.6)          & 96.5 (94.3-98.0)          \\
Residents, model-quality images    & 79.6 (72.9-85.3)          & 97.7 (96.0-98.8)          & 82.4 (74.1-88.8)          & 92.4 (89.8-94.5)          & 33.3 (19.7-49.5)          & 88.7 (85.8-91.1)          \\
Residents, full-quality images     & 81.5 (75.0-86.9)          & 99.1 (97.8-99.7)          & 95.1 (89.6-98.1)          & 89.6 (86.6-92.0)          & 41.7 (26.7-57.9)          & 90.2 (87.6-92.5)          \\
Residents, model-aided performance & 97.5 (94.2-99.2)          & 100.0 (99.4-100.0)        & 96.1 (90.9-98.7)          & 96.0 (94.0-97.5)          & 58.3 (42.1-73.3)          & 92.6 (90.2-94.5)          \\
\end{tabular}}
\end{table}

\begin{figure}
  \centering
  \includegraphics{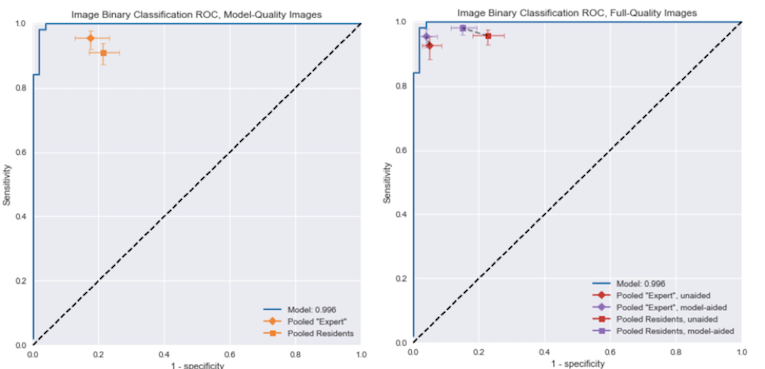}
  \caption{model ROC vs human observers. Image on left shows model’s ROC vs sensitivity and specificity with 95\% CIs for the human observers when using model-quality images. On the right is shown model’s ROC vs these metrics when human observers use full-quality images in both unaided and aided conditions. Note that this only reflects performance in binary fracture detection task and does not reflect performance in subclassification task.}
  \label{fig:fig9}
\end{figure}

Comparisons of relevant binary and multiclass Cohen’s kappa coefficients are shown with significance values in Table 8. When human observers evaluated “model-quality” images, the model achieved statistically significant superior performance in both binary and multiclass tasks. When human observers used the full-quality images, the model tended to achieve superior performance, but only with statistical significance over the resident group. Experts achieved significantly superior performance when they evaluated full-quality rather than model-quality images. When using the model as an aid to human performance, experts reached essentially equivalent performance to the model, while residents continued to be outperformed regarding binary classification. When used as an aid to human observers, the model tended to improve all humans’ performance, although this reached statistical significance only for residents. Interestingly, while experts achieved superior performance relative to residents under all conditions tested, aided resident performance did not differ significantly from unaided expert performance.

\begin{table}[]
\centering
\caption{difference in Cohen’s kappa values with significance values calculated using randomization test with 10000 permutations (bolded if p \textless 0.05)}
\label{tab:table8}
\begin{tabular}{l|ll}
\hline
                                                                                                                       & \begin{tabular}[c]{@{}l@{}}Binary Cohen’s \\ kappa difference\\ (p-value)\end{tabular} & \begin{tabular}[c]{@{}l@{}}Multiclass Cohen’s\\ kappa difference\\ (p-value)\end{tabular} \\
                                                                                                                       \hline
\begin{tabular}[c]{@{}l@{}}Model-quality images,\\ model vs experts (one-tailed)\end{tabular}                          & \textbf{.140 (.0097)}                                                                  & \textbf{.128 (.0061)}                                                                     \\
\begin{tabular}[c]{@{}l@{}}Model-quality images,\\ model vs residents (one-tailed)\end{tabular}                        & \textbf{.223 (.0001)}                                                                  & 0.2183 (\textless{}.0001)                                                                 \\
\begin{tabular}[c]{@{}l@{}}Full-quality images,\\ model vs experts (one-tailed)\end{tabular}                           & .045 (.1458)                                                                           & .040 (.1784)                                                                              \\
\begin{tabular}[c]{@{}l@{}}Full-quality images,\\ model vs residents (one-tailed)\end{tabular}                         & \textbf{.190 (.0014)}                                                                  & \textbf{.180 (.0005)}                                                                     \\
\begin{tabular}[c]{@{}l@{}}Model-aided performance,\\ model vs experts (one-tailed)\end{tabular}                       & .005 (.3838)                                                                           & \textbf{.001 (.4831)}                                                                     \\
\begin{tabular}[c]{@{}l@{}}Model-aided performance,\\ model vs residents (one-tailed)\end{tabular}                     & .090 (.0467)                                                                           & .057 (.0972)                                                                              \\
\begin{tabular}[c]{@{}l@{}}Experts, full-quality vs \\ model-quality images (one-tailed)\end{tabular}                  & .095 (.0092)                                                                           & .088 (.0033)                                                                              \\
\begin{tabular}[c]{@{}l@{}}Experts, model-aided vs \\ unaided performance (one-tailed)\end{tabular}                    & \textbf{.040 (.1048)}                                                                  & \textbf{.039 (.0744)}                                                                     \\
\begin{tabular}[c]{@{}l@{}}Residents, full-quality vs \\ model-quality images (one-tailed)\end{tabular}                & \textbf{.033 (.2089)}                                                                  & \textbf{.039 (.1126)}                                                                     \\
\begin{tabular}[c]{@{}l@{}}Residents, model-aided vs \\ unaided performance (one-tailed)\end{tabular}                  & .100 (.0025)                                                                           & .123 (\textless{}.0001)                                                                   \\
\begin{tabular}[c]{@{}l@{}}Experts vs residents, \\ model-quality images\end{tabular}                                  & .083 (.0550)                                                                           & .0905 (.0092)                                                                             \\
\begin{tabular}[c]{@{}l@{}}Experts vs residents, \\ both unaided with full-quality\\  images (two-tailed)\end{tabular} & .145 (\textless{}.0001)                                                                & .140 (\textless{}.0001)                                                                   \\
\begin{tabular}[c]{@{}l@{}}Unaided-experts vs \\ aided-residents (two-tailed)\end{tabular}                             & \textbf{.045 (.2024)}                                                                  & \textbf{.017 (.523)}                                                                      \\
\begin{tabular}[c]{@{}l@{}}Experts vs residents, \\ both aided (two-tailed)\end{tabular}                               & .085 (.0087)                                                                           & .056 (.0326)                                                                             
\end{tabular}
\end{table}

\section{Discussion}
In this study we demonstrate at least expert-level binary and multiclass classification of hip radiographs into one of six categories in both fractured and non-fractured groups. To our knowledge this represents the first report of fracture subclassification by deep learning in the literature. Our excellent results are notable given the limited size of our training set, which was only 1849 images, which we believe we overcame with aggressive use of data augmentation, and the validity of our ground truth. As we labeled radiographs, we referred to subsequent imaging, including CT, MRI and post-surgical radiographs whenever the classification was not obvious. Dominguez et al showed that up to 10\% of hip fractures are occult on radiographs (24); therefore, solely using radiographs as ground truth may lead to substantial avoidable bias due to misclassification. However, because of the potential morbidity of missing a diagnosis of hip fracture, patients with negative radiographs and high clinical suspicion for hip fracture (hip pain after fall, inability to ambulate, etc.) often undergo advanced imaging with CT or MRI, which serves as a more reliable ground truth than plain radiographs (25). Additionally, as the functional classification of hip fracture dictates the type of operation that a patient receives, the final surgery choice serves as a reliable ground truth for multiclass classification.

In our comparison to fellowship-trained experts, our model showed statistically superior performance when experts used model quality images, and a non-statistically significant trend towards superior performance when experts used the full-quality images. Using human expert performance as a proxy for Bayes’ optimal error rate, this demonstrates that few gains are likely to be made in our system using the low-resolution images via further hyperparameter optimization, and therefore, efforts should be focused rather on developing a model that can process higher resolution images. This notion is validated by the statistically significant boost in expert performance between the lower and full quality images, indicating that some information essential to classification may be lost in down-sampling, and that we may improve our model’s performance if trained on larger resolution images. In this project we were restricted to using low resolution given our small dataset size and need for ImageNet-pretraining; future research will explore boosting our training set size in a self-supervised fashion using natural language processing and the automated hip detector described in this paper which we hope will allow us to escape the resolution constraints of using an ImageNet-pretrained model.

As fellowship-trained radiologists and orthopaedists are often not the persons responsible for reading hip radiographs in the emergency room, we included senior residents in emergency medicine, orthopaedics, and radiology in our comparison to human performance. The model achieved statistically superior performance to residents when using both model-quality and full resolution images. Additionally, we showed that when using the model as an aid the residents improved their performance, approximating the level of fellowship-trained experts. These results together suggest a model such as ours may be used to decrease diagnostic error and reduce use of advanced imaging in the emergency room.

These results build on a growing body of evidence that suggest the clinical utility of deep learning in musculoskeletal radiography. Lindsey et al recently showed excellent results of a modified u-net architecture in detection of wrist fracture on radiograph, and similar to the present study showed a significant boost in human performance when given the model’s predictions as an aid (16). Regarding hip fractures, Gale et al demonstrated radiologist-level performance of binary classification by comparing the model’s performance to the radiologist reports (26), and Urakawa et al demonstrated orthopaedist-level detection of intertrochanteric fractures when using model-quality images (27). To our knowledge no prior study has performed subclassification of fracture types.

\subsection{Limitations}
\begin{figure}
  \centering
  \includegraphics{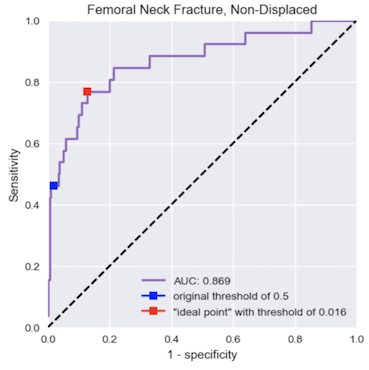}
  \caption{Receiver-operator curve showing change in sensitivity and specificity of non-displaced femoral neck fracture detection by varying the detection threshold.  With original model threshold of 0.5, sensitivity for this type of fracture is 46.2\% (95\% CI, 28.2-64.9) with specificity of 97.8\% (95\% CI, 96.1-98.9). After setting the threshold to reach the curve’s ideal point, which is the point on the line minimizing distance from the top-left corner, the sensitivity is 77.0\% (95\% CI 58.5-90.0\%) with specificity of 87.4\% (95\% CI, 83.9-90.3\%).}
  \label{fig:fig10}
\end{figure}
The limitations of this study include the fact that all of our radiographs come from one institution, potentially limiting its generalizability, although we mitigated this by using 20 years of images obtained with many different scanners. An additional limitation is that the classification algorithm depends on a bounded box image, which was generated manually. To this end we trained the object detection algorithm described above and demonstrate here equivalent classification performance with these automatically-generated boxes, demonstrating a fully-automated  end-to-end solution with deep learning.

Another limitation in this study is that our model only considers a single image in its prediction, unlike a human interpreter, who may look at several views. For example, apparently subtle femoral neck fractures are often best seen on the lateral image, which was not included in our model. Rayan et al recently demonstrated excellent results from a novel system that used a CNN as a feature-extractor for images in a given radiographic study and then fed this output into a recurrent neural network to generate study-level predictions for pediatric elbow fractures (29). Such a system may help to improve our model’s performance and represents an exciting area of research.

The largest limitation of the model presented is the relatively low sensitivity to nondisplaced femoral neck fractures, with only 58\% correctly identified as a fracture of some kind in the test set and only 46\% correctly subclassified. These are challenging fractures to diagnose, as shown in Table 7, which demonstrates that human observers performed even more poorly than the model under all conditions for this fracture subtype. As these are often subtle, we believe that increasing the image resolution and including multiple views into the model’s prediction may improve performance, and we are actively exploring these directions as described above. Interestingly, the model does a relatively high-performing ROC for this fracture subtype with AUC 0.869, but as shown in figure 10, the prediction threshold of 0.5 results in operating far from the ideal point on this specific curve. If we adjust the detection threshold to reach the ideal point (the point that minimizes the distance from the top-left of the figure), multiclass sensitivity improves to 77.0\% (95\% CI 58.5-90.0\%) with specificity of 87.4\% (95\% CI, 83.9-90.3\%). This suggests a role for the model suggesting further imaging with CT or MRI if it’s predicted likelihood of non-displaced femoral neck fracture lies above this ideal point’s threshold even though the most likely prediction is no fracture. Such a model applied on this test set would suggest additional imaging on just 6.5\% of patients yet would lead to 80.8\% of these fractures being correctly identified as a fracture of some kind.

\subsection{Conclusion}
Hip fractures are a common cause of morbidity and mortality globally, and recent literature suggests that early operative stabilization of hip fractures is essential to optimize outcomes. This study demonstrates at least expert-level performance of automatic hip fracture diagnosis by a fully-automated end-to-end deep learning-based system, with functional subclassification that allows stratification into operative groups. Additionally, we demonstrate that when used as a diagnostic aid our model improves human performance, with aided residents approximating the performance of unaided fellowship-trained experts. Such a system has the potential to decrease diagnostic error and the use of advanced imaging, while improving outcomes by decreasing the time to surgery, which may have a significant impact on patient recovery and morbidity.

\bibliographystyle{unsrt}  


\end{document}